\documentclass[preprint]{aastex631}

\newcommand{\vela}{Vela\,X-1}

\usepackage{amsmath}

\begin{document}

\title{A NICER viewing angle on the accretion stream of Vela X-1}

\author{Roi Rahin}
\affiliation{NASA Goddard Space Flight Center, 8800 Greenbelt Road, Greenbelt, MD 20771, USA}
\affiliation{Physics Department, Technion, Haifa 32000, Israel}
\author{Ehud Behar}
\affiliation{Physics Department, Technion, Haifa 32000, Israel}



\begin{abstract}

Vela X-1 is the archetypical eclipsing high-mass X-ray binary, composed of a neutron star (NS) accreting the B-star wind.
It was observed by nearly all X-ray observatories, often multiple times, 
featuring a rich spectrum of variable emission lines.
Yet, the precise origin of these lines in the binary system remains uncertain. 
We perform a systematic, orbital phase-dependent analysis of the reflected Fe\,K$\alpha$ fluorescence line at 6.4\,keV using over 100 NICER observations.
We resolve the line variability into 500\,s time bins and find that it is predominantly due to variation in the ionizing flux, 
with a moderate underlying phase dependence over the 9\,day orbital period.
Our analysis reveals a significant reflection component that cannot originate from the companion B-star alone.
We also find that an appreciable portion of the B-star surface is obscured opposite the eclipse, and this obscuration is not symmetric around the mid-point (phase=0.5).
We argue that an accretion stream, from the B-star to the NS and distorted by the orbital motion, is responsible for both the additional fluorescence emission component and for obscuring the B-star.
\end{abstract}



\section{Introduction} \label{sec:intro}

Vela X-1 is a high mass, eclipsing, X-ray Binary (XRB). Vela X-1 consists of a neutron star (NS) of $\sim2M_\odot$ and a massive B-star ($M_*\sim25 M_\odot$,$R_*\sim30R_\odot$) with a short orbital period of $\sim9$\,days \citep{kreykenbohm2008high}. The system is highly compact, with a separation of only $\sim1.8R_*$, which implies the NS is deeply embedded in the dense stellar wind. Vela X-1 is a consistent X-ray source of moderate luminosity ($\sim 10^{36}$\,erg\,s$^{-1}$), but at a short distance of $\sim2$\,kpc \citep{kretschmar2021revisiting}, which makes it one of the brightest X-ray point sources in the sky.
The X-ray spectrum of Vela X-1 is commonly modeled with an absorbed cut-off (above 20\,keV) power-law continuum, photo-ionized emission line components, and a reflection component with both continuum and (neutral) fluorescence lines.

Vela X-1 is highly variable in X-ray luminosity, line of sight column density, and emission line spectrum. The luminosity variations are attributed to giant flares, off-states, and column density variations \citep[and references therein]{kreykenbohm2008high,kretschmar2021revisiting}. 
The presence of an accretion stream from the B-star enshrouding the NS is evidenced by partial absorption of the 2-20\,keV X-ray continuum, but not the harder X-rays between 15-50\,keV, even when the NS is in front of the B-star at phase 0.5 (see Figure 4 in \citet{kretschmar2021revisiting}).
The variability in column density is associated with the accretion stream and clumps in the stellar wind \citep{watanabe2006x,grinberg2017clumpy}. Simulations of high-mass XRBs show this column density is expected to be asymmetric around phase 0.5 \citep{blondin1991enhanced,manousakis2012neutron}, with observations in good agreement \citep{kretschmar2021revisiting}. The emission line spectrum is variable as well. 
\citet{grinberg2017clumpy} noted extreme variability in ionized line emission around phase 0.25. 
The reflection component and its fluorescence lines vary as well and are assumed to originate in a cold and dense neutral medium. Previous works ascribed most of the fluorescence to the surface of the stellar companion. High observed equivalent width (EW) of the Fe\,K$\alpha$ line at phases 0.5 and 0.25 implied an additional reflection component \citep{watanabe2006x}. 

The cause of the line variability is a standing question in \vela\ and may contribute to understanding the geometry of the binary and its accretion stream. Furthermore, the origin and physical mechanism of both ionized emission and fluorescence in \vela\ is still debated \citep{2008AIPC.1054....3M,kretschmar2021revisiting}. A systematic analysis of multiple phases and accretion states may improve our understanding of this enigmatic source. 
The challenge is to disentangle the slowly varying phase dependence from that of the stochastic accretion process.
In this paper, we harness the high sensitivity and high cadence observations of \vela\ by NICER (NS Interior Composition Explorer) to do just that.

\section{Vela X-1 observations and spectral models}

\subsection{NICER observations and modeling}
\label{sec:NICER}
NICER \citep{gendreau2016neutron} is a soft X-ray (0.2 – 12 keV) telescope mounted on the ISS (International Space Station). NICER observes Vela X-1 regularly as a calibration source, but many of these observations are valid for science analysis.
We use all observations of Vela X-1 up to 2022-05-04 and divide them into segments of up to 500 continuous seconds using Ftools\footnote{\texttt{\detokenize{http://heasarc.gsfc.nasa.gov/ftools}}}. In practice, some segments contain less than 500 seconds of data after applying NICER's Good-Time-Intervals procedure. 
We use a pyXspec\footnote{\texttt{\detokenize{https://heasarc.gsfc.nasa.gov/xanadu/xspec/python/html/index.html}}} script to analyze all resulting observation segments that have $>10000$ total photons in the 3-9\,keV band. 
This excludes data around the eclipse, from phases 0.9 - 0.1.
We use a model combining an absorbed power-law, a gaussian emission line for the 6.4\,keV Fe-line, and unabsorbed continuum reflection  \citep{magdziarz1995angle}. The final Xspec model used by the fitting script is:

\begin{equation}
    TBabs*(TBabs*powerlaw + pexrav + gauss)
\end{equation}

\noindent The first TBabs absorption component is the Galactic absorption derived from HI maps \footnote{\texttt{\detokenize{https://heasarc.gsfc.nasa.gov/cgi-bin/Tools/w3nh/w3nh.pl}}} and its column density is set to $0.37\times10^{22}$ cm$^{-2}$. 
The second TBabs absorption component is intrinsic to Vela X-1 and is fitted for each NICER segment. 
The model assumes the absorbing medium is a neutral cold gas with solar composition.  
We originally attempted to use TBfeo to fit the Fe abundance, but the Fe abundance can not be well constrained using this data, and we thus returned to solar abundances.
The power-law slope and norm are fitted, and both are tied to the pexrav power-law slope and norm. In the pexrav model, the abundance is set to the solar value, while the angle cosine, defined between the line of sight and reflector inclination, is set to 0.95 since the inclination of Vela X-1 is known to be close to $90^{\circ}$, and the dominant reflection is expected to be the B-star surface. 
The power-law cutoff energy is set to 20\,keV \citep{makishima1999cyclotron}. Since the X-ray source is a compact object, it is easily obscured by wind clumps and accreting matter. By comparison, the reflector is at least the size of the B-star, with a radius of $\sim30R_\odot$, and will only be obscured by large-scale geometrical elements. Therefore, we remove the pexrav direct power-law component (by assigning the rel\_refl component a negative value), to allow the power-law component to be absorbed separately while keeping the reflection component unabsorbed.
In summary, the free model parameters are the power-law slope and norm, the column density of the local absorber, the gaussian emission line parameters, and the reflection scaling.

\subsection{Chandra and XMM-Newton observations}

We utilized Chandra/HETG observations, processed by The Chandra Grating-Data Archive and Catalog, TGCat \citep{huenemoerder2011tgcat}, and XMM-Newton/RGS observations to measure emission lines from Vela X-1. The eight Chandra/HETG observations of Vela X-1 were analyzed multiple times \citep[e.g.,][]{schulz2001ionized,goldstein2004variation,watanabe2006x,grinberg2017clumpy,rahin2020canonical}. The three RGS observations were analyzed by \citet{martinez2014accretion}. For a more comprehensive list of observations see \citet{kretschmar2021revisiting}. In this work, We use the Chandra/HETG observations 1927,1926,19953,19952 and 14654 as well as the XMM-newton observation 0841890201 (see Table\,\ref{tab:obs}). 

\vspace{0.2cm}
Throughout this work, We calculate the orbital phase of each measurement using the reported eclipse epoch and orbit period from  \citet{kreykenbohm2008high}.
Note that phases calculated in this way may show a slight deviation from other works.

\section{sporadic vs. phase-dependent spectral variability}
As a binary system, Vela X-1 is expected to exhibit periodic, phase-dependent behavior. 
Such behavior can be seen in the averaged observed X-ray flux over the 9\,d period, although surprisingly the peak flux in the 2-20\,keV range is found around phase $\sim0.3$ \citep{kretschmar2021revisiting}. 
On the other hand, Vela X-1 varies on much shorter time scales, likely due to the sporadic accretion process.
Spectral elements such as column density, line flux, and power-law slope show substantial variability on the shortest time scales where variability can be decidedly measured ($\sim$1\,h). 
In this section, we try to identify and separate these two forms of variability, phase-dependent and sporadic, by focusing on the column density and the Fe\,K$\alpha$ line flux, with a note about the photo-ionized gas.

\subsection{Column density and complete obstruction}

\begin{figure} [!htbp]
\begin{center}
\includegraphics[scale=0.35]{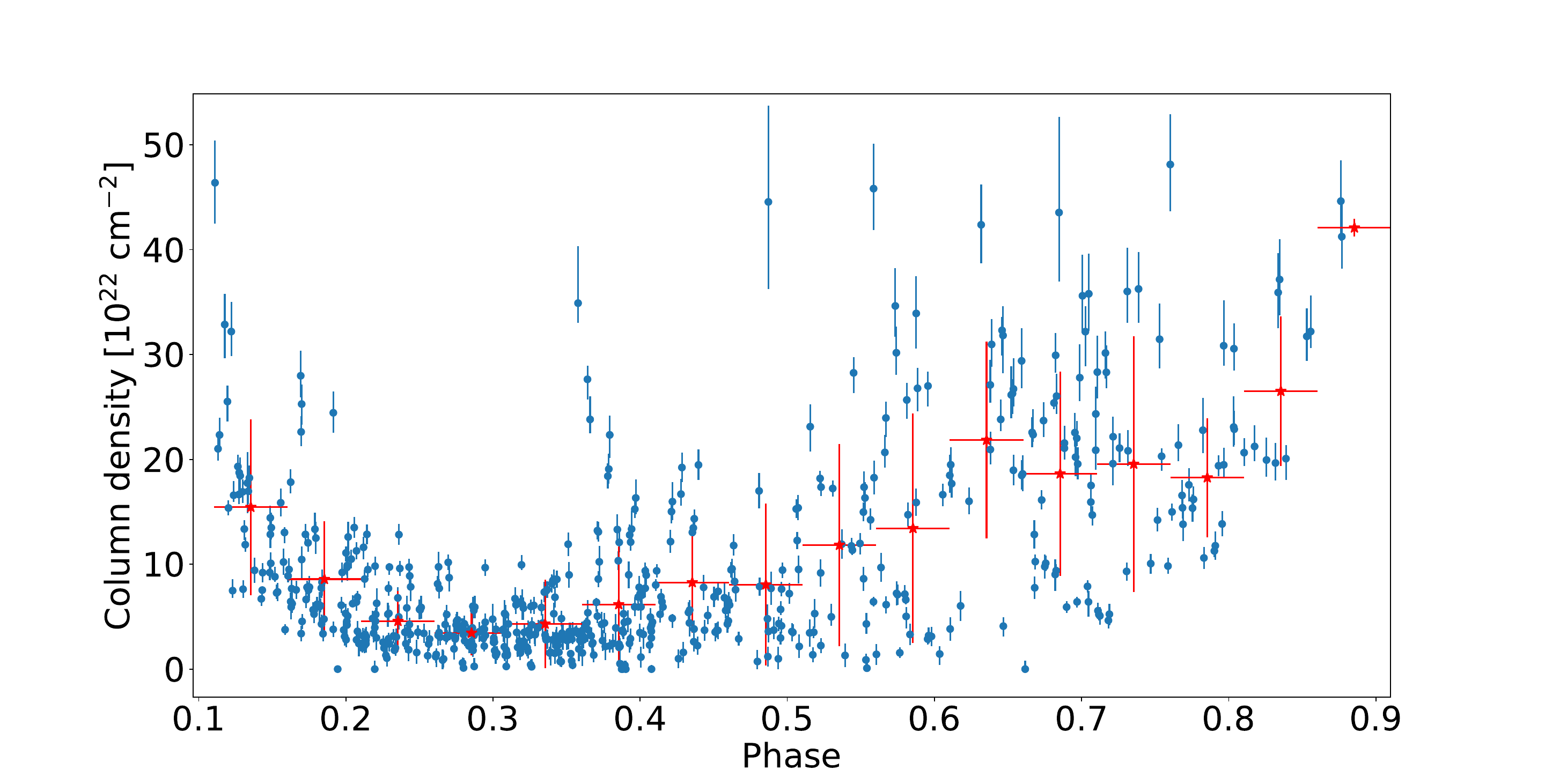}
\end{center}
\caption 
{ \label{fig:NH_phase} Equivalent $N_\textrm{H}$ measured in various NICER observation slices of up to 500\,s. Blue dots are individual measurements and red stars are averages over 0.05 phase bins. Note that the red error bars represent the standard deviation between measurements, not the error of the mean.
A significant trend is visible where $N_\textrm{H}$ decreases to a minimum at phase $\sim\!0.3$ and then increases. 
This asymmetric behavior with phase indicates excess gas on one side of the system, which is likely associated with a (rotating) accretion stream.
}
\end{figure}

The column density absorbing the X-ray continuum of \vela\ is known for high variability both with phase and between observations \citep{kretschmar2021revisiting}. Figure\,\ref{fig:NH_phase} shows the column density across 608 NICER observation segments of Vela\,X-1. Each observation is at most 500\,s long.
Blue dots represent individual measurements and red stars are averages over 0.05 phase bins. The red error bars represent the standard deviation between different measurements. When performing averages, we first average all observations that are temporally adjacent (defined as being at most one hour apart) and then take the average of all other observations in the same bin.
Since Vela X-1 is known to have periods of extreme absorption \citep{liao2020spectral}, we exclude outliers with $N_\textrm{H} > 50\times 10^{22}$\,cm$^{-2}$ from both the figure and the averaging process. This excludes 22 measurements both for figure clarity and to avoid skewing the average with extremely high outlier measurements. Almost all excluded measurements lie between phases 0.3 and 0.6.
 Both the low number of excluded measurements and their much higher column density lends support to their classification as outliers.
The X-ray source is more absorbed on average at phases 0.5--0.9 than at 0.1--0.5 indicating a breaking of symmetry in the system, where high column density gas is located (and rotates with the system) on one side of the binary axis.
This effect, suggesting an accretion stream as the absorber, was previously noted by others \citep[and references therein]{kretschmar2021revisiting} and fits high-mass XRB simulations \cite{blondin1991enhanced,manousakis2012neutron}, but it is shown here with significantly more data points, owing to NICER's sensitivity and numerous observations.

\subsection{The Fe K$\alpha$ line}

One of the most conspicuous features of the Vela X-1 X-ray spectrum is the Fe K$\alpha$ fluorescence line. This line was reported in many observations and studied extensively \cite{nagase1994line,goldstein2004variation,watanabe2006x,grinberg2017clumpy,kretschmar2021revisiting}, and yet there is no clear understanding of the line source or its phase-dependence. 
We use the 500\,s sliced NICER observations (Section\,\ref{sec:NICER}) to obtain 630 measurements of the Fe K$\alpha$ line. 
Although the actual variability time of the line may be even shorter than 500\,s, to the best of our knowledge this is the largest sample with such short times segments to be presented for \vela.
We use it to shed light on the physical origin of this line.

\begin{figure} [!htbp]
\begin{center}
\includegraphics[scale=0.35]{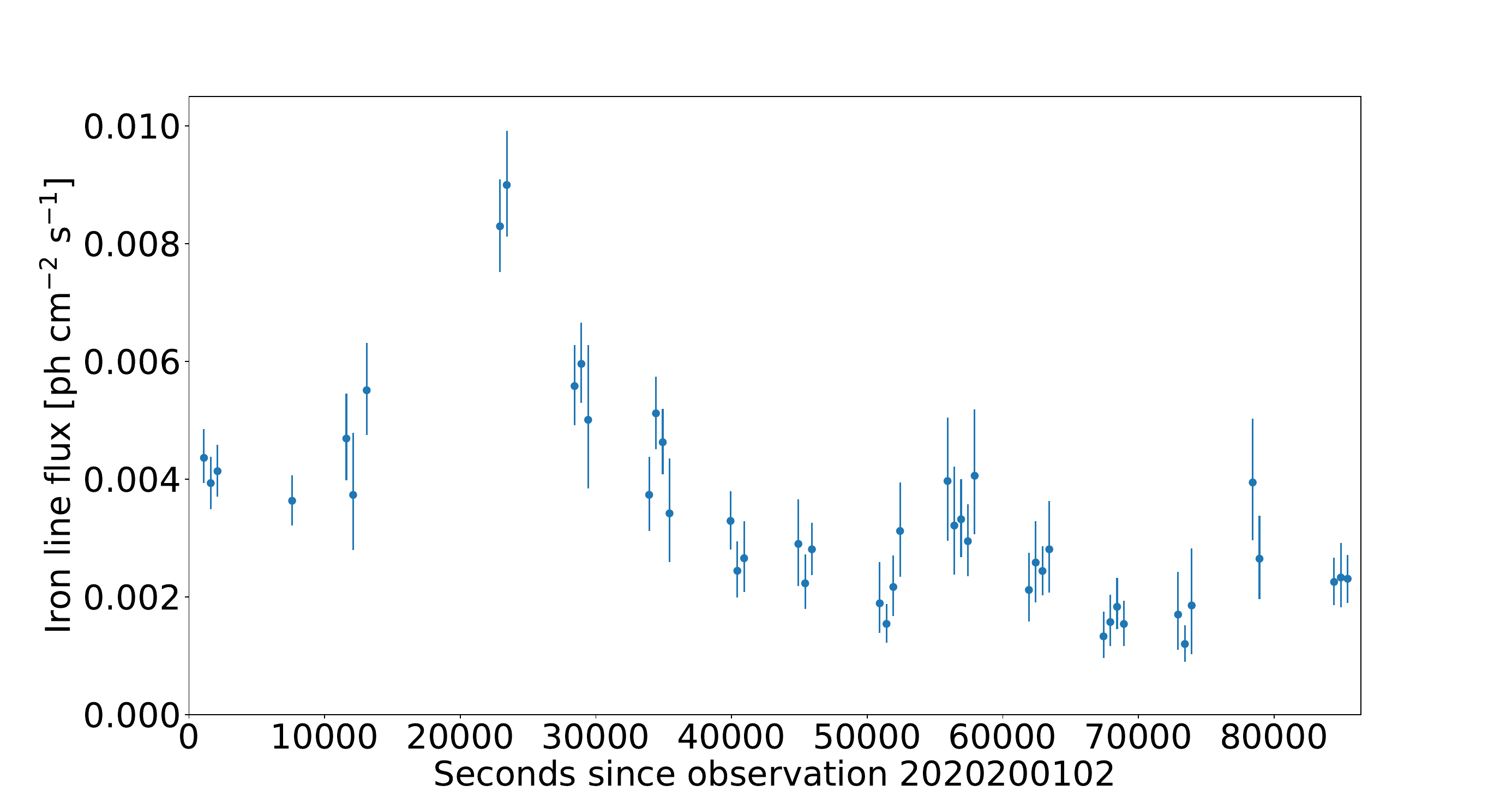}
\end{center}
\caption 
{ \label{fig:variability} The Fe K$\alpha$ line flux variations over 1 day as recorded with several NICER observations. Each observation was broken into segments of at most 500\,s. There is a significant variation of the line flux, sometimes on a timescale of one hour.}
\end{figure} 

\begin{figure} [!htbp]
\begin{center}
\includegraphics[scale=0.35]{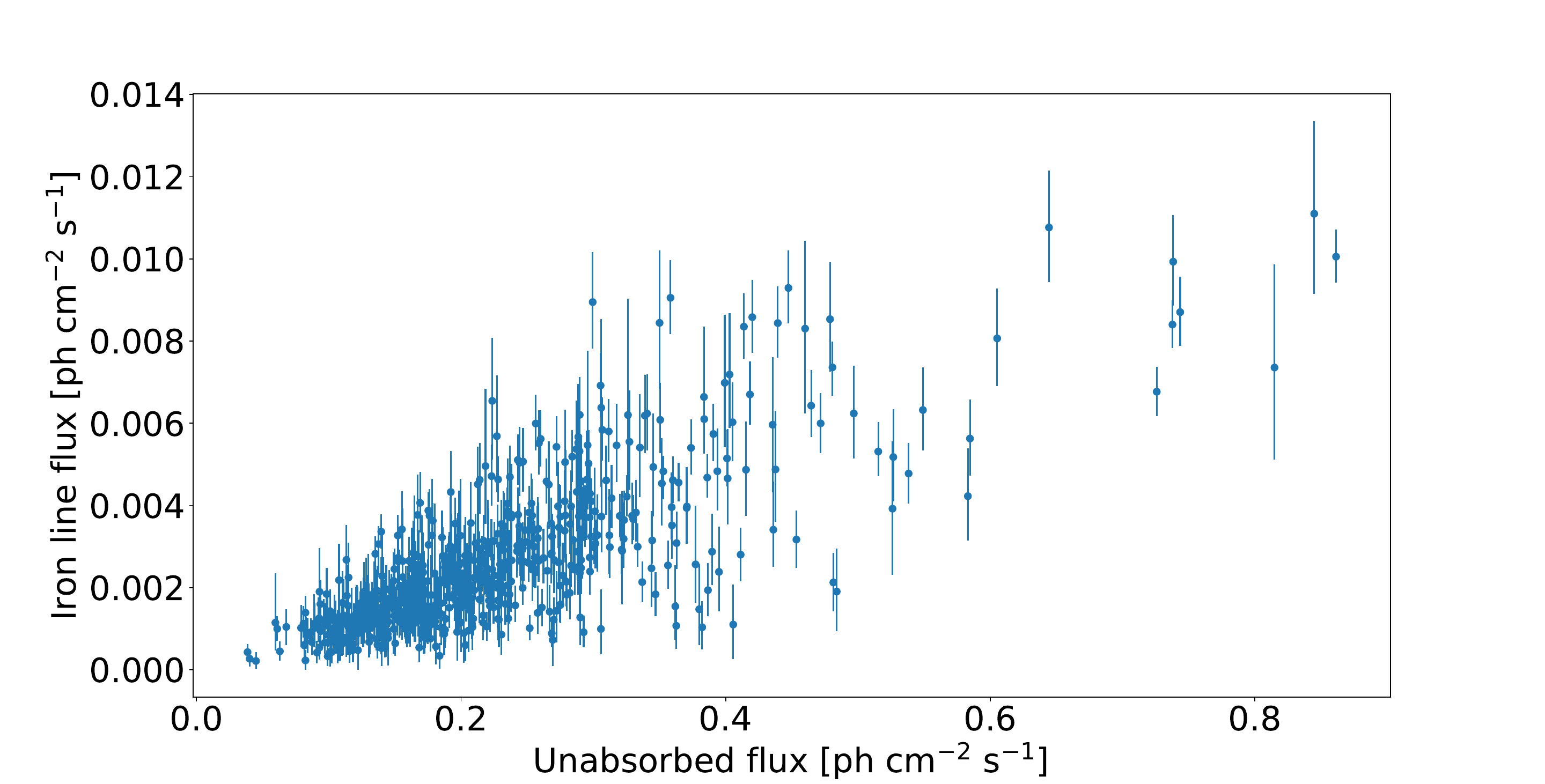}
\end{center}
\caption 
{ \label{fig:unabs} The Fe K$\alpha$ line flux vs. the unabsorbed continuum flux in the 7-20 keV range. The correlation is readily apparent and expected.}
\end{figure}

Figure\,\ref{fig:variability} shows Fe K$\alpha$ flux measurements from the start of observation 2020200102 over a duration of 1 day. The measurements are broken into 500\,s segments and show the rapid variability of the line. 
The figure shows 
significant difference up to a factor of 2 in line flux between observations ($\sim5$\,ks). 
Variability between adjacent 500\,s segments seems to exist but is not significant to the level that we can measure here.
Figure \ref{fig:unabs} shows the relation between the Fe\,K$\alpha$ flux and the model unabsorbed flux in the 7\,keV to 20\,keV range derived from the spectral model described in Section\,\ref{sec:NICER}. 
The Pearson correlation coefficient is $0.78$ with $p<0.001$. 
Since we expect the line to be primarily driven by the \vela\ continuum, this relation is as expected.

\begin{figure} [!htbp]
\begin{center}
\includegraphics[scale=0.35]{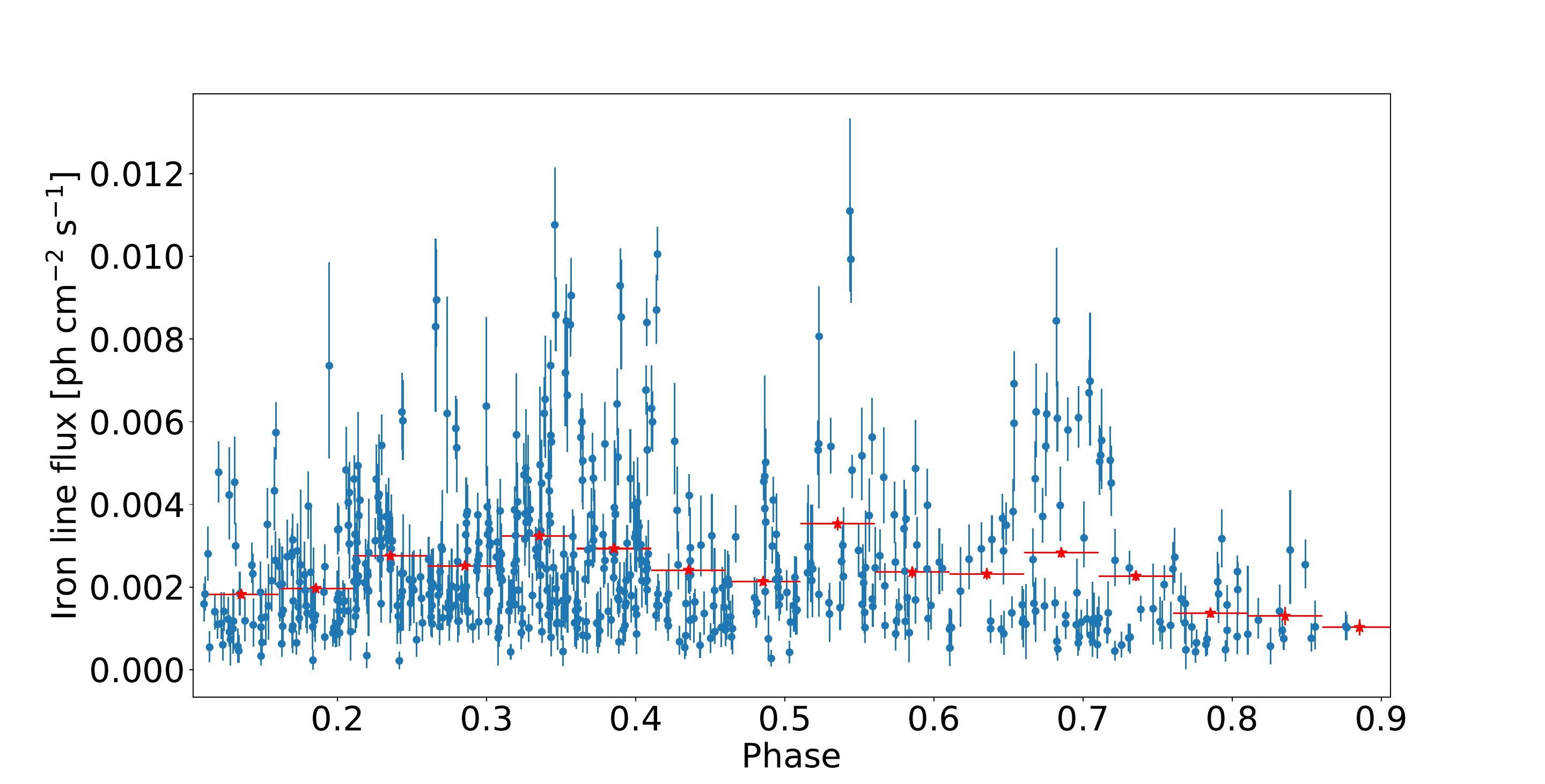}
\end{center}
\caption 
{ \label{fig:phase_fe} Fe K$\alpha$ line flux with phase. Red stars are average values over bins of size 0.05. No clear dependence on phase emerges.
}
\end{figure}

The above time scales are much shorter than the orbital period. We wish to also understand the effect of the orbital phase on the 
Fe K$\alpha$ line. 
Figure\,\ref{fig:phase_fe} shows the Fe K$\alpha$ line flux across multiple measurements in various phases. In an attempt to detect trends, We calculated the average flux over bins of size 0.05. 
Unlike the correlation with continuum flux, no discernible pattern emerges, and the mean line flux remains between 0.0015--0.004 photons\,cm$^{-2}$\,s$^{-1}$. 

\begin{figure} [!htbp]
\begin{center}
\includegraphics[scale=0.35]{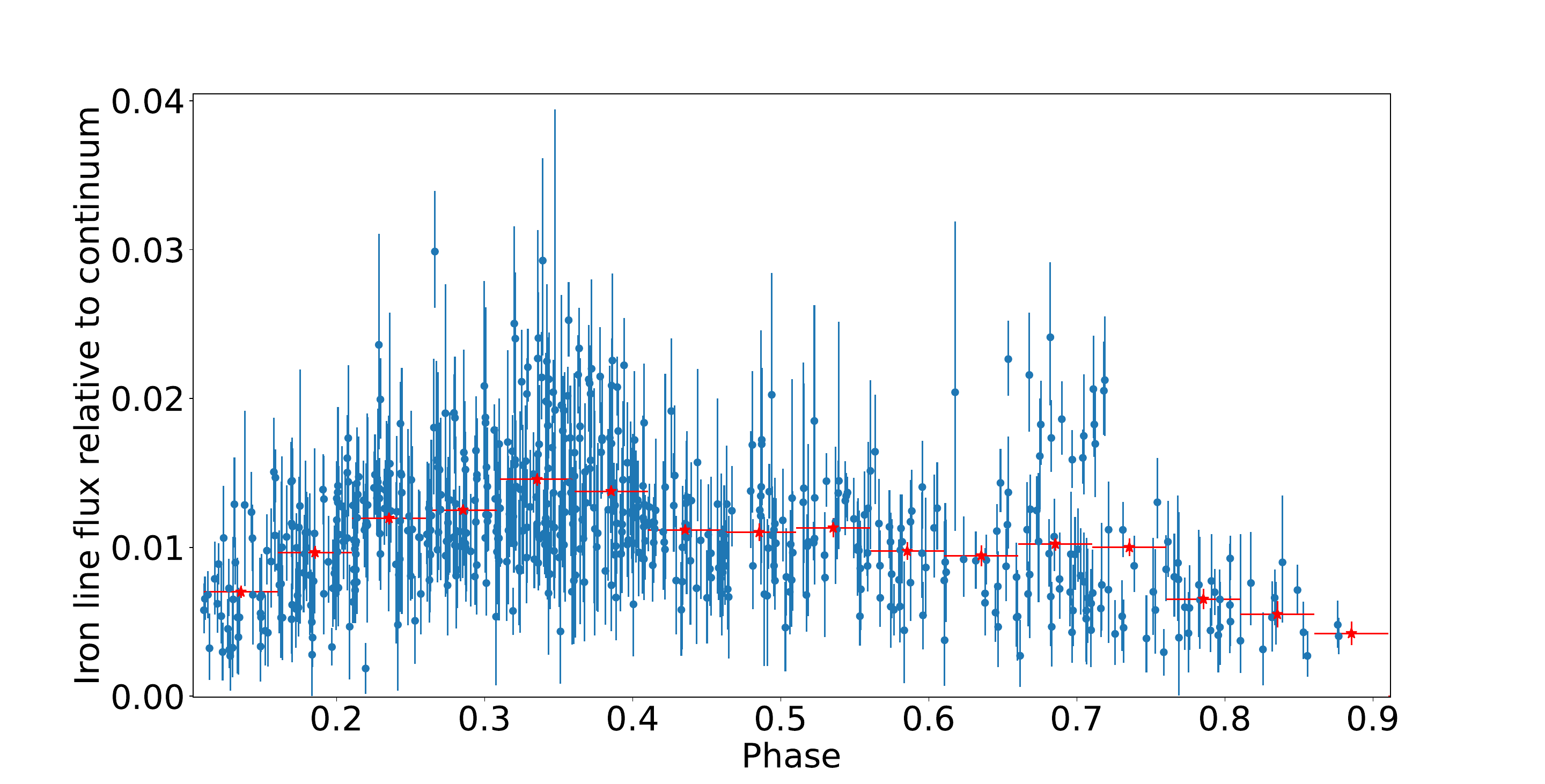}
\end{center}
\caption 
{ \label{fig:phase_fe_unabs} The Fe K$\alpha$ flux relative to the continuum flux in the 7-20 keV range with phase. The red stars are average values over bins of size 0.05. There is a clear increase in relative flux up to phase $\sim 0.35$ followed by a slow decrease. The shape is clearly asymmetrical.}
\end{figure}

Since we aim to discern the phase dependency of the Fe\,K$\alpha$ line, we must disentangle the strong dependence on the continuum. We thus divide the line flux by the ionizing unabsorbed model flux in the $7-20$\,keV range.
Figure\,\ref{fig:phase_fe_unabs} shows this fraction of the iron line relative to the continuum with phase, revealing a trend that was not visible in Figure\,\ref{fig:variability}. 
Once the continuum effect is folded out, this trend can shed light on the geometry of the binary. 
In Figure\,\ref{fig:phase_fe_unabs}, we see a rise in relative flux with phase up to $\sim$0.35, which interestingly is also where the $2-20$\,keV continuum peaks \cite{kretschmar2021revisiting}. 

Furthermore, in and out of the eclipse the line flux relative to the continuum shows an increase towards a virtual maximum at the center (Figure\,\ref{fig:phase_fe_unabs}). 
This is expected if the reflector is the half-sphere surface of the B-star, rotating into and out of our line of sight, analogous to the lunar phases. 
However, two strong deviations from this behavior are apparent. 
First, in the middle of the period between phases $0.3 -0.7$ the relative line flux is much lower than expected from this trend, implying that part of the reflecting star is obstructed from our line of sight. 
The second deviation is that the line flux relative to the continuum does not tend to zero towards the eclipse (at phases 0.9 and 0.1), as would be expected from a half-lit rotating sphere. 
On the contrary, there is significant Fe\,K$\alpha$ emission, of the order of $0.5\%$ relative to the continuum, throughout the orbit. 
Note that although some line emission is present around the eclipse in phases $0.9 - 0.1$, there is no continuum to divide by, and no data is available for Figure\,\ref{fig:phase_fe_unabs}.

\subsection{Simple geometrical reflection model}
\begin{figure} [!htbp]
\begin{center}
\includegraphics[width=0.5\linewidth]{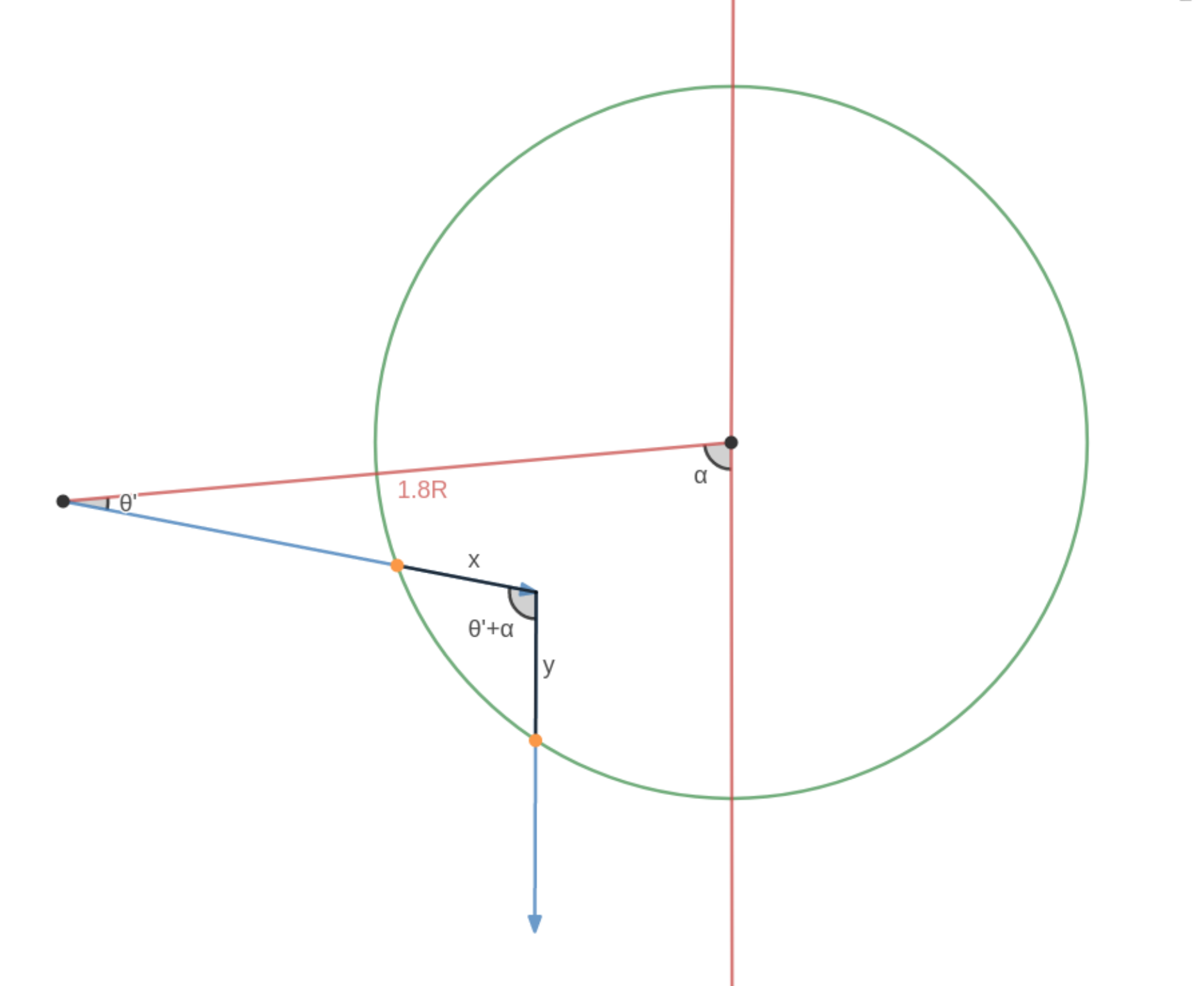}

\end{center}
\caption
{ \label{fig:sphere_model} The 2D model for 
reflection of a point source (the NS at the leftmost point) from an optically thick sphere, into our line of sight (the observer is at the bottom reflecting an 0.25 or 0.75 phase), drawn with Desmos
(\texttt{\detokenize{https://www.desmos.com/geometry}})
.
The stellar surface is denoted by the green circle. 
The ionizing continuum enters and exits the sphere through a path length of $x$ and $y$, respectively.
}
\end{figure}

The Fe K$\alpha$ line is associated with the reflection of the X-rays from the direction of the NS to our line of sight, which is highly dependent on geometry.
Since the primary suspect as a reflector is the B-star, we first attempt a model of reflection from an optically thick sphere, 
similar to the initial assumption of \citet{watanabe2006x}. 
The present calculation generalizes the 1D slab approximation described in \citet{rahin2020canonical}.
Unlike in a slab, the entry (continuum excitation) and exit (of the fluorescence photon) paths are different.
The exit path length to our line of sight is calculated geometrically for all incident angles entering the sphere, and further depends on the entry path length, and the orbital phase. 
The model includes the quantum probability to excite a Fe\,K$\alpha$ line inside the star, and its probability to escape the medium \citep[see][]{rahin2020canonical}.
It ignores Compton scattering, whose effect is smaller than the measurement uncertainties \citep{george1991x}. See Figure\,\ref{fig:sphere_model} for a 2D drawing of the geometry. A detailed derivation of the reflected line intensity can be found in Appendix \ref{sec:geometry}.

\begin{figure} [!htbp]
\begin{center}
\includegraphics[width=1\linewidth]{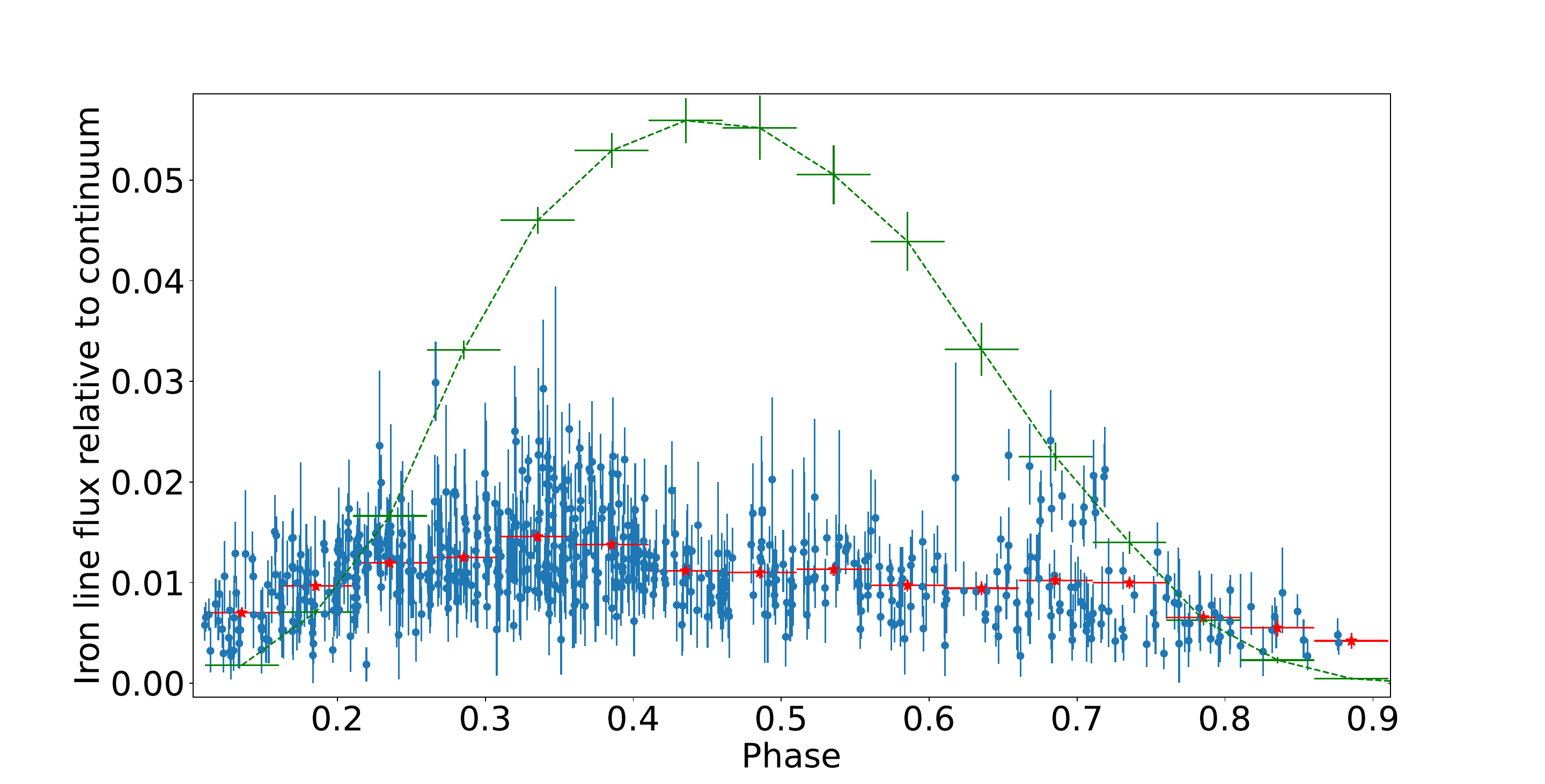}

\end{center}
\caption 
{ \label{fig:phase_fe_sphere} Relative Fe K$\alpha$ line flux with phase (blue, and 0.05 phase averaged in red), compared to the expected reflection from the surface of a sphere (green) relative to the continuum. 
The model over-predicts the measurements between phase $0.2-0.7$ and under-predicts it elsewhere.
Clearly, reflection from a sphere alone cannot explain the observed line flux.}
\end{figure}

Figure\,\ref{fig:phase_fe_sphere} shows the Fe K$\alpha$ flux and the theoretical value from the spherical reflection model.  As can be seen, the model over-predicts the reflected line flux around phase 0.5 and under-predicts it close to the eclipse. Specifically, the average flux measured around the 0.5 phase is about 5 times lower than expected and the average flux measured near eclipse (phase 0.1-0.15) is about 5 times higher than expected. The over-prediction could be a result of one or more factors. The star could be partially obscured, either from us or from the source itself. In the latter case, the ionizing radiation exciting the material on the face of the B-star would be lower than assumed by the model, lowering the expected emission. The ionizing radiation could also have a different continuum spectrum than that used in the model. For example, if the power-law photon index of the radiation reflected off the star is higher than the measured power-law index, the expected fluorescence will be considerably lower.
Compared to the model over-prediction, the under-prediction is more difficult to explain within the confines of the model. The model predicts little emission close to the eclipse, but the observed emission is still considerable. Any attempt to reconcile this discrepancy within the sphere-reflection model will necessarily only increase the model over-prediction. 
It must therefore be deduced that the model describing reflection from a sphere cannot describe the observed Fe K$\alpha$ flux or its orbital dependence. This implies both the presence of additional reflecting material apparent most clearly near the eclipse and the presence of material obscuring a large portion of the surface of the star, most conspicuously between phases $0.3-0.7$. 

\pagebreak
\subsection{Photoionized-gas emission lines}
\label{sec:ionized}
\begin{figure} [!htbp]
\begin{center}
\includegraphics[scale=0.35]{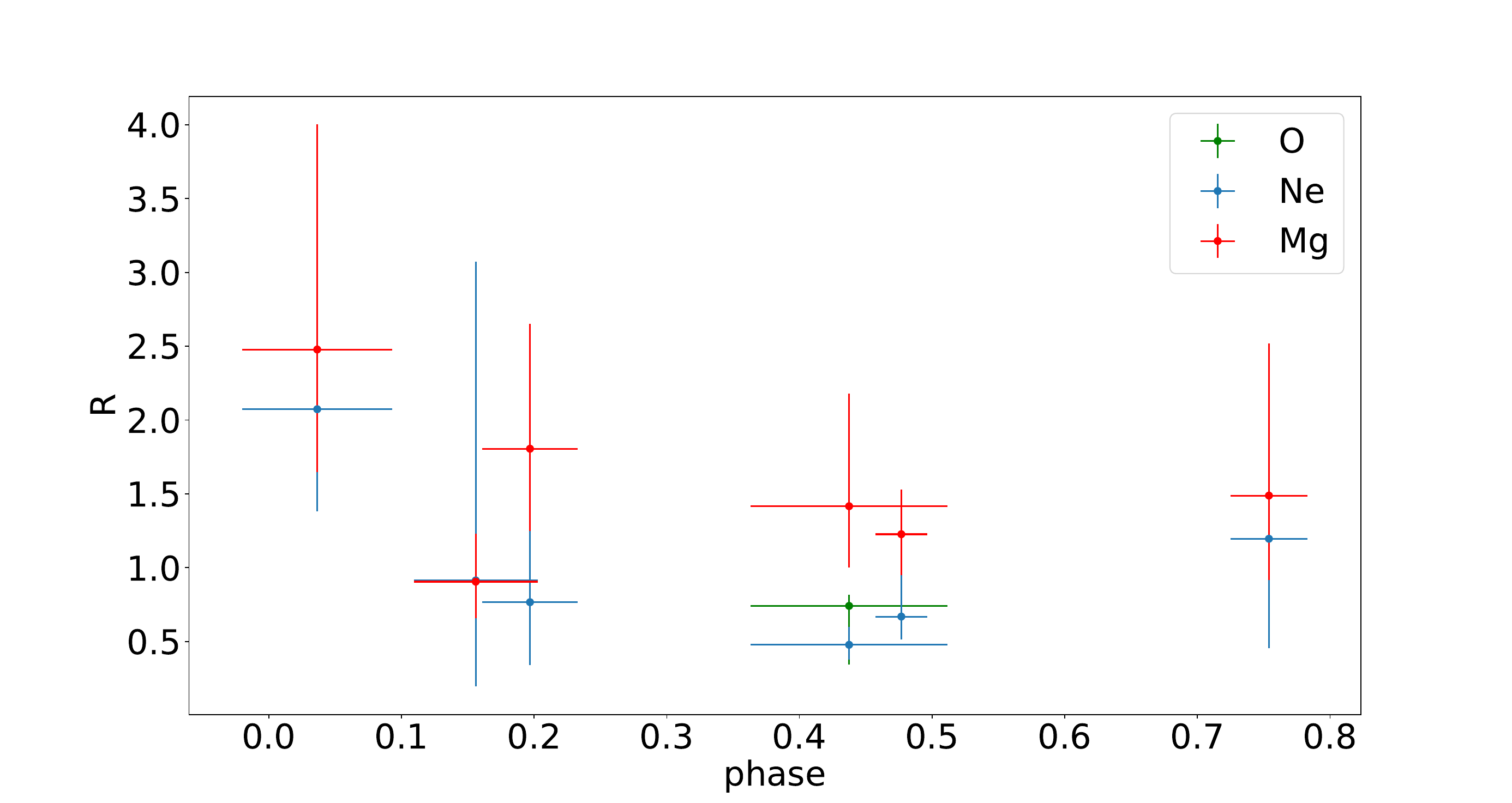}
\end{center}
\caption 
{\label{fig:R_all} The $R=f/i$ ratio of O, Ne, and Mg across several orbital phases of Vela X-1. 
Phase 0.43 measurements are from the RGS, and include O, while all other measurements are from HETG (see Appendix\,\ref{App_obs}). Measured $R$ values are much lower than the nominal ones of 3.7, 3.1, and 2.7 for O, Ne, and Mg respectively, implying a significant effect of high UV flux or electron density. }
\end{figure}

The Vela X-1 Chandra/HETG and XMM-Newton/RGS spectra are rich in emission lines from a variety of ion species. 
Here, we rely on the He-like ions to provide insights into the properties of the emitting gas, in an attempt to complete the geometrical picture of \vela . 
Specifically, the $R=f/i$ ratio of the forbidden ($f$) to intercombination ($i$) lines can probe the electron density and distance from the B-star \citep{porquet2000x,leutenegger2006measurements}. 
Both high electron number density and high B-star UV flux incident on the X-ray gas will reduce the $R$ ratio from its nominal low-density value.
Figure\,\ref{fig:R_all} shows the $R$ ratio of He-like O, Ne, and Mg at several orbital phases. 
Some phase dependency is suggested with $R$ having higher values during and around the eclipse.

\begin{figure} [!htbp]
\begin{center}
\includegraphics[scale=0.35]{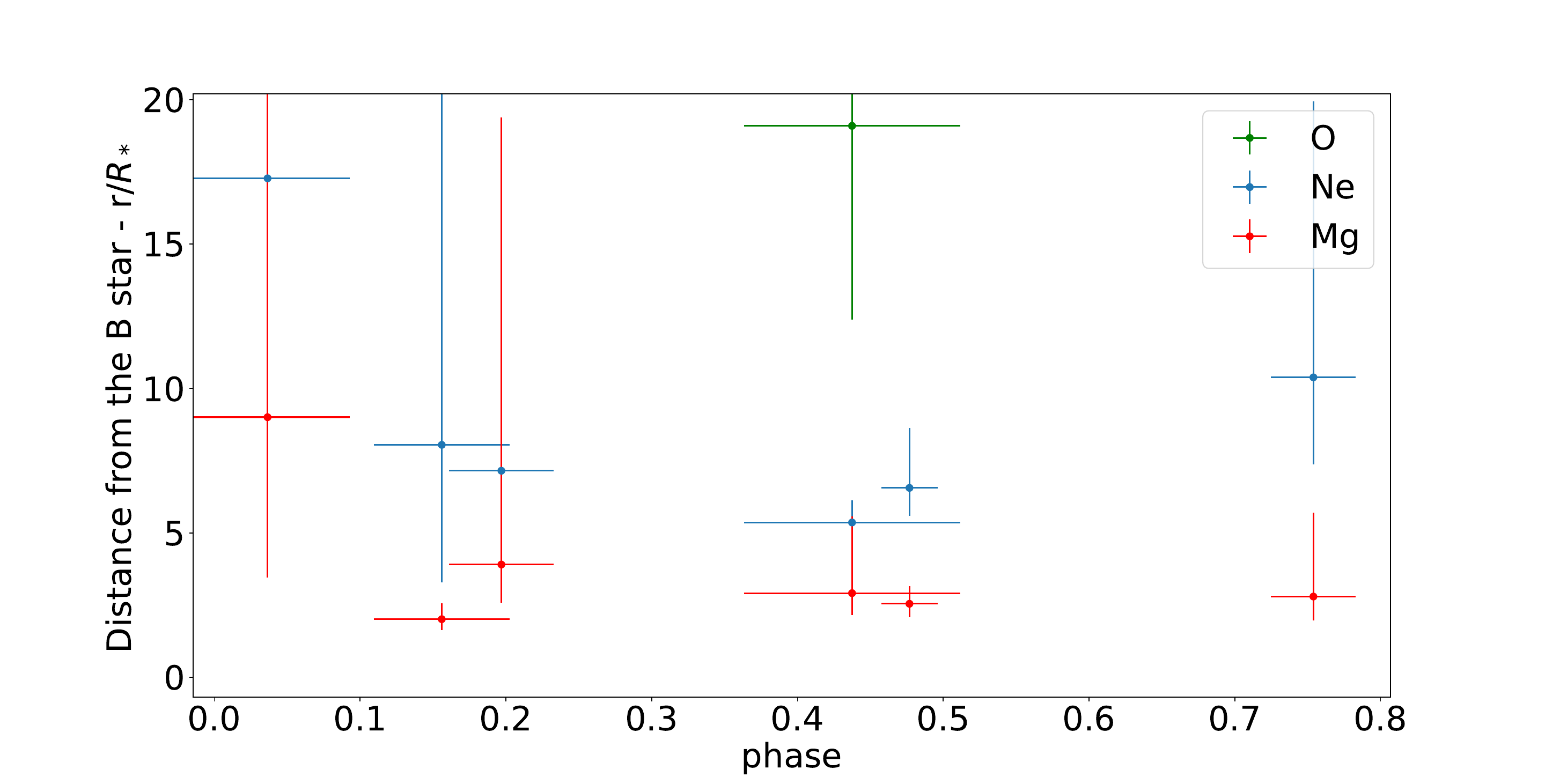}
\end{center}
\caption 
{ \label{fig:r_star} Distance of He-like O, Ne, and Mg ions from the OB donor star across several phases of \vela. The distance is calculated from the UV effect on the R-ratio at the low-density limit. Notice the consistently shorter distance with atomic number, with Mg the closest to the B-star, and O the farthest at $\sim20R_*$. }
\end{figure}

Differentiating UV flux and electron density effects in photo-ionized plasma can be challenging.
They decrease together with the distance from the B-star (both as $r^{-2}$), implying the effects of both on the R ratio increase and decrease in tandem. This was previously noted by \citet{peretz2019direct} in AGN observations. 
We will now argue that in the case of Vela X-1, the dominant effect is UV pumping.
Since the donor in the \vela\ binary is a B-star, it is natural to first consider the impact of the UV radiation field.
Figure\,\ref{fig:r_star} shows the implied distances from the B-star in stellar radii $r/R_*$, given the measured $R$ ratios, as a function of phase. 
We rely on the method introduced by \citet[][Eqs.\,1-4 and Table\,1 therein]{leutenegger2006measurements}, and adopt the stellar parameters used by \citet{grinberg2017clumpy} ($T_{eff} = 25,000$ K, $\log g = 3.0$), who performed a similar calculation for the Chandra observation 1928.
We find that the He-like lines are emitted from $2-20\,R_*$ from the UV source, presumably the B-star.
The distance of the NS from the B-star is $\sim1.8R_*$.
This implies the emitting gas is farther than the NS, sometimes considerably. 
For Mg, the difference is small and may be a result of model assumptions, but for Ne and O, the distance of the emitting gas is $r>5R_*$. 
The emitting Oxygen gas is particularly far at a distance $r>12R_*$ from the B-star. 

In theory, lower than nominal $R$ values can also be caused by a high number-density of the emitting gas. The densities for which the $R$ ratio of O, Ne, and Mg deviates from its nominal value are greater than $10^9$\,cm$^{-3}$, $10^{10}$\,cm$^{-3}$, and $10^{11}$\,cm$^{-3}$, respectively \citep{porquet2000x}. 
We can obtain a rough approximation for the stellar wind density profile by assuming for simplicity a density profile $n(r) = n_0 / (r/R_*)^2$. 
The total (radial) column density of such a wind is $N_\textrm{H} = n_0 R_*$. 
For $N_\textrm{H} = 3-4\times10^{22}$\,cm$^{-2}$ \citep{rahin2020canonical} and $R_* = 2\times10^{12}$\,cm, $n_0 \sim 10^{10}$\,cm$^{-3}$. This estimated density is consistent with those obtained in simulations \citep{2008AIPC.1054....3M,manousakis2011accretion}.
Since the wind density decreases rapidly from this value with $r$, the only element that might be affected by it is O.
However, O was already shown to originate far from the B-star at $r/R_* > 10$, where $n/n_0 < 100$ 
(Figs.\,\ref{fig:R_all}, \ref{fig:r_star}).
The effect of gas density on the $R$ ratio in a smooth wind is therefore negligible. It is still possible that the He-like lines originate from dense wind clumps, but these would have to be even farther away from the star than indicated in Fig.\,\ref{fig:r_star}. 

We did not include the $R$ ratio of He-like Si in this discussion, despite it being well measured with values between $3.5-8$, since this is substantially {\textit higher} than the theoretical nominal limit of 2.3 \citep{porquet2000x}. This anomaly was noted by \citet{grinberg2017clumpy} and by \citet{amato2021looking}, with no explanation. Unfortunately, this work cannot offer a definitive explanation either. 
Examinations of other photo-ionized X-ray sources, such as the Circinus Galaxy with $R\textrm{(Si)}\sim4$, hint that 
the high value is prevalent. The peculiarly high R ratio can have several possible explanations. 
First, there could be an instrumental artifact in the Chandra/HETG which is not corrected for by the response functions. 
Given the close proximity of the intercombination and forbidden lines, this is unlikely. 
Second, if a 1s electron is ionized from the ground state of Li-Like Si, the resulting He-like ion is in the 1s2s excited state.
This will contribute to the emission in the forbidden line and increase R.
However, the fraction of Li-like ions in steady-state is small.
Third, and most likely, the theoretical values taken from \citet{porquet2000x} may require updating. 
Future observations of Vela\,X-1, and other X-ray sources with XRISM/Resolve \citep{tashiro2020status} may aid in determining the exact cause of these high R ratios for Si, which may call for re-examining the theoretical values.

The upshot of this analysis of the He-like line ratios is that the ionized gas is located several stellar radii away from the B-star and is likely unrelated to the reflecting medium discussed above.
\section{Discussion}

The column density observed throughout the orbit of Vela X-1 shows a clear asymmetry around phase 0.5, with low $N_\textrm{H}$ in phases 0.1--0.5 and high  $N_\textrm{H}$ in phases 0.5--0.9. These results agree with other measurements \citep[and references therein]{kretschmar2021revisiting} and with simulations of eclipsing high-mass XRBs \citep{blondin1991enhanced,manousakis2012neutron}. 
As explained in these references, this is a result of the Coriolis force pushing away the accretion stream to one side of the system.
Comparing our results to the simulations of \citet{manousakis2012neutron} (Figures 2 and 3 therein), we note the lack of a column density "bump" around phases 0.4--0.6. Considering the estimated mass of the NS in \vela\ of $1.7-2.2\,M_\odot$, the best matching parameters in their simulations would be $v_\infty \sim 1200$ km\,s$^{-1}$ and a wind mass-loss rate of $\dot{M}_W \sim 10^{-6} M_\odot$\,yr$^{-1}$. These rough estimates are in agreement with previous observations \citep{nagase1986circumstellar,prinja1990terminal}, but here derived from the phase dependence of the column density.

The fluorescence spectrum of \vela, considering its high intensity, must originate as a reflection from a dense, cold region with a high column density. 
A simple model for reflection from the spherical surface of the B-star overestimates the observed flux between phases 0.3 and 0.7 by up to a factor of 5 but underestimates the observed flux around the eclipse by a similar factor (Figure\,\ref{fig:phase_fe_sphere}). These drastic discrepancies are unlikely to be fixed by small modifications to the model. 
We propose that an accretion stream in \vela\ is responsible for both of these deviations.
The fact that the reflected line relative to the continuum is not symmetric with phase (Figure\,\ref{fig:phase_fe_unabs}) indicates the stream is located on one side of the symmetry axis of the binary.
It obscures part of the B-star between phases 0.35--0.75 and adds steady fluorescence emission, $\sim 0.5\%$ of the continuum, throughout the orbit.
This interpretation requires the accretion stream to be dense enough to have a low ionization parameter and to emit Fe\,K$\alpha$. On the other hand, it can not be Compton thick, since in phases 0.5--0.9 we observe the reflection through the stream. 

The majority of the reflection component was postulated to originate on the surface of the companion star \citep{watanabe2006x}. However, with the stellar wind, this surface may not be well defined and other components such as wind clumps, and accretion streams may contribute to the reflection components.
Indeed, \citet{watanabe2006x} observed with Chandra excess EW at two discrete phases 0.25 and 0.5 compared to the reflection contribution of the B-star and its wind, in their model. 
They thus invoked a reflecting cloud to explain this excess. 
The present analysis of the line flux relative to the unabsorbed continuum above 7\,keV is more physical since it is this continuum that excites the line, rather than the local continuum at 6.4\,keV that does not.
In fact, we find a \textit{deficiency} of relative flux at phase 0.5 (Figure\,\ref{fig:phase_fe_unabs}). 
NICER's full orbital coverage and shorter observation times presented in this work enable a more complete picture of the reflection with phase, and thus consistently point to the \vela\ accretion stream.
Note that using the EW for this kind of analysis could be misleading since variable continuum absorption, for instance, will change the EW without any change in line flux.
It appears that the emission from the accretion stream is visible from both sides of the binary (phases 0.25 and 0.75), as the deviation from a reflecting sphere appears on both ends of the phase plot. 

Further verifying this hypothesis can be done using high-resolution spectroscopy of Vela X-1 immediately before and after the eclipse. 
The fluorescence line ratios of different elements can then be compared to estimate the fluorescing medium column density \citep{rahin2020canonical}.
Immediately after the eclipse, the sole reflecting medium that can be observed is the accretion stream, and this measurement will yield its column density. Similarly, immediately before the eclipse, the accretion stream is observed from the other side and a measurement of the fluorescing column density viewed through the stream should yield similar values.
XRISM, expected to launch in 2023, is a suitable instrument for such an observation.

As shown in Section\,\ref{sec:ionized}, the emission lines of the ionized gas in \vela\ provide information about its location. 
Our analysis of the He-like $R$ ratios and their dependence on UV flux indicates that the minimum distance of the He-like Ne and Mg ions from the B-star is greater than the distance to the NS. This result was reported by \citet{grinberg2017clumpy} for phase 0.25, but we show it to be consistent across all phases.
Hence, the ionized gas is more remote than the putative accretion stream. 
In hydrodynamical simulations by \citet{2008AIPC.1054....3M}, He-like Si is found between the NS and B-star with the majority of emission in a region close to the surface of the B-star $(r \approx 1R_*)$. 
Since He-like Mg forms in similar $\xi$ values as He-like Si, our findings of $r = 2R_*$ do not agree with these results. 
The emission of ionized Ne and Mg during the eclipse (phase\,0) appears to originate farther from the star than between phases 0.1--0.9. 
This is consistent with the overall weakness of the ionized lines in eclipse. Indeed, \citet{watanabe2006x} concluded that the majority of ionized lines are produced in a region obscured by the host B-star during the eclipse.

Interestingly, in our results the ions appear ordered in distance by decreasing atomic numbers, with Mg originating closest to the star, Ne following, and O originating some distance away, at almost $20R_*$ (Figure\,\ref{fig:r_star}). 
This could imply a gradient in the ionization parameter of the stellar wind.
The emission of He-like O specifically, probably originates in the sparse wind of the B-star illuminated by the X-ray source. Taken together, these results imply a co-rotating ionization cone of gradually decreasing ionization diverging away from the binary system (increasing $r$). This implies the density decreases less steeply than $n \propto r^{-2}$, suggesting collimation, possibly by magnetic fields as found by \citet{fukumura17} for the black hole XRB GRO\,J1655+40.  
One could test this hypothesis by systematically measuring line fluxes at various phases, especially near eclipse. 
If the line emission truly originates farther away from the binary system, the fluxes and $R$ ratios of the He-like triplets should be (near) constant throughout the orbit, including during the eclipse. Testing this hypothesis will be possible with XRISM/Resolve observations of \vela\ during the eclipse and in phases close to eclipse.
We caution that the varying nature of the X-ray continuum of \vela\ is likely to affect ionized line fluxes as well. Thus, a number of observations at each phase is required to obtain a mean distance before meaningful conclusions can be drawn, as done here for Fe\,K$\alpha$.

\section{Conclusions}

From the results presented in this work, the following conclusions can be drawn:

\begin{itemize}
    \item{NICER's complete coverage in phase space of \vela\ allowed us to separately monitor the continuum flux, the fluorescence lines, and the column density as a function of phase.}
    \item{The observed Fe\,K$\alpha$ line flux relative to the hard continuum between $7-20$\,keV, when compared with a simple geometrical model deviates from a smooth orbital reflection model from the B-star, and points to a substantial role of an accretion stream in both obscuring the B-star and reflecting X-rays from the NS source into our line of sight.}
    \item{The UV-sensitive line ratios of He-like ions from Chandra/HETG and XMM-Newton/RGS indicate distance estimates of several stellar radii from the B-star.}
    \item{A decreasing ionization with distance from the binary suggests collimation of the ionization cone, perhaps by magnetic fields.}
\end{itemize}
\begin{acknowledgments}
R.R. is supported by an appointment to the NASA Postdoctoral Program at the NASA Goddard Space Flight Center, through a contract with ORAU. R.R. is supported in part by the Zuckerman STEM Leadership Program. R.R. was supported by a Ramon scholarship from the Israeli Ministry of Science and Technology. E.B. acknowledges support from a Center of Excellence of THE ISRAEL SCIENCE FOUNDATION (grant No. 2752/19). 

This work is based in part on observations obtained with NICER.
The scientific results reported in this article are based in part on observational data obtained from the Chandra Gratings Catalog and Archive.
This work is based in part on observations obtained with XMM-Newton, an ESA science mission with instruments and contributions directly funded by ESA Member States and NASA. 

\end{acknowledgments}

\vspace{5mm}
\facilities{NICER \citep{gendreau2016neutron}, Chandra/HETG \citep{canizares2005chandra}, XMM-Newton/RGS \citep{den2001reflection} }

\software{Xspec \citep{arnaud1996astronomical}, pyXspec, HEAsoft \citep{heasarc2014heasoft}}

\bibliography{bibliography}{}
\bibliographystyle{aasjournal}

\appendix

\section{Spherical reflection model}
\label{sec:geometry}
Assuming a binary inclination of $90^\circ$, the exit path length from a circle of radius $R$ when assuming the X-ray source (origin) is at a distance of $1.8R$ from the center is:

\begin{equation}
\begin{aligned}
    &y = -1.8 \cos (\alpha)+\cos (\alpha+\theta') \left(x+1.8-\sqrt{1-3.24 \sin ^2(\theta')}\right) + \\ 
    & + \bigg(3.6 x \cos (\theta')-x^2 -2 x \left(1.8\, -\sqrt{1-3.24 \sin ^2(\theta')}\right) \\
    &\left(\cos (\alpha+\theta') \left(-\sqrt{1-3.24 \sin ^2(\theta')}+x+1.8\right)-1.8 \cos (\alpha)\right)^2 \bigg)^{0.5}
\end{aligned}
\end{equation}

Where $x$ is the length of the path into the sphere, $\alpha$ is the angle between the line of sight and the line connecting the center of the sphere and X-ray source, and $\theta'$ is the angle of incidence of the entry path. All lengths are in units of $R$. Figure\,\ref{fig:sphere_model} is the drawing this 2D model describes.

Since the B-star itself has an average number density in excess of $10^{16}$ cm$^{-3}$, we can assume $x \ll R$ and calculate $y(x)$ to first order, $y=Cx+O\left(x^2\right)$:

\begin{equation}
\begin{aligned}
& C_{2D} = \frac{\left(1\, -0.556 \sqrt{1-3.24 \sin ^2(\theta')}\right) \cos ^2(\alpha+ \theta')- \cos (\alpha) \cos (\alpha+ \theta')}{\left(0.556 \sqrt{1-3.24 \sin ^2(\theta')}-1\right) \cos (\alpha+ \theta')+\cos (\alpha)} +\\
&+ \frac{0.556 \sqrt{1-3.24 \sin ^2(\theta')}+ \cos (\theta')-1}{\left(0.556 \sqrt{1-3.24 \sin ^2(\theta')}-1\right) \cos (\alpha+ \theta')+\cos (\alpha)}+\cos (\alpha+ \theta')
\end{aligned}
\end{equation}

Where $C_{2D}$ refers to the solution for the 2D problem. 

We can now consider the same problem in 3 dimensions. If we assume the inclination is $90^\circ$, the exiting photons always travel on a plane parallel to the one in Figure\,\ref{fig:sphere_model}. Therefore, all we need to do is transform $\theta'$ to $\theta$ and $\phi$. This adds the following 2 corrections:
\begin{equation}
\theta' = \arctan(\frac{\sin(\theta)\cos(\phi)}{\cos(\theta)})
\end{equation}
\begin{equation}
 C(\theta,\phi) = C_{2D} \cdot \sqrt{\cos(\theta)^2 + \sin(\theta)^2\cos(\phi)^2}
\end{equation}

where $\theta$ and $\phi$ are the standard spherical coordinates with the z-axis being the line connecting the NS and the center of the B-star.

The resulting expected flux is therefore:
\begin{equation} \label{eq:thick_sphere}
F = A_{\textrm{Z}} \omega_{K\alpha} \int\displaylimits_{E_{K,Z}}^{\infty} \left( \int \frac{I_0 (\frac{E}{E_0})^{-\Gamma} \sigma_{Z,K}(E)}{\sigma(E)+C(\theta,\phi)\sigma(E_{K\alpha,Z})}d\Omega \right)dE 
\end{equation}

Where $\Omega$ covers the visible sphere cap from the point-of-view of the X-ray source. The other parameters are as described in \citet{rahin2020canonical}.

\pagebreak
\section{Observations and measurements}
\label{App_obs}

\begin{deluxetable*}{c|c|c|c|c}[ht]
	\tablecaption{List of high-resolution observations used}
	\label{tab:obs}

	\tablehead{\colhead{\textbf{Instrument}} & \colhead{\textbf{Observation ID}} &\colhead{\textbf{Date observed}} & \colhead{\textbf{Exposure (s)}} & \colhead{\textbf{Phases}}}
	\startdata
	Chandra HETG & 1926 & 2001-02-11 & 83150 & 0.98 - 0.09 \\
	Chandra HETG & 1927 & 2001-02-07 & 29430 & 0.46-0.5  \\
	Chandra HETG & 14654 & 2013-07-30 & 45880 & 0.72-0.78  \\
	Chandra HETG & 19953 & 2017-02-22 & 70470 & 0.11-0.2  \\
    Chandra HETG & 19952 & 2017-06-19 & 54330 & 0.16-0.23  \\
    XMM-Newton RGS &0841890201 & 2019-05-03& 113500 & 0.36-0.51 \\	\enddata
	
\end{deluxetable*}

\begin{deluxetable*}{c|c|c|c|c|c}[ht]
	\label{tab:vela_R}
	\tablecaption{R measurements for various Vela X-1 phases}
	\tablehead{\colhead{\textbf{Observation ID}} & \colhead{\textbf{Phase}} & \textbf{Ne R)} &\textbf{mg R}&\textbf{O R}&\textbf{Si R}}
	\startdata
	1926 & 0.98 - 0.09 & $2.07_{-0.69}^{+1.52}$&$2.48_{-0.83}^{+1.52}$& -&$4.57_{-1.46}^{+2.67}$\\
	1927 & 0.46-0.5 &$0.67_{-0.15}^{+0.34}$ &$1.23_{-0.28}^{+0.30}$ &- &$3.66_{-0.73}^{+1.0}$\\
    14654 & 0.72-0.78 &$1.28_{-0.48}^{+0.99} $&$1.36_{-0.48}^{+0.84}$ & - &$6.98_{-2.4}^{+5.93}$\\
	19953 & 0.11-0.2  & $0.91_{-0.72}^{+2.16}$&$0.91_{-0.25}^{+0.33}$ & -&$12.54_{-6.24}^{+49.87}$\\
    19952 & 0.16-0.23 & $0.77_{-0.43}^{+0.71}$&$1.81_{-0.56}^{+0.85}$ & -&$5.8_{-1.4}^{+2.26}$\\
    0841890201 & 0.36-0.51 & $0.48_{-0.1}^{+0.12}$& $1.42_{-0.42}^{+0.76}$ &$0.74_{-0.4}^{+0.07}$ &-\\
	\enddata
	
\end{deluxetable*}

\end{document}